\documentclass[letterpaper, 10 pt, conference]{ieeeconf}  

\IEEEoverridecommandlockouts                              

\usepackage{graphicx} 

\usepackage{amssymb,amsmath,amsfonts}
\usepackage[UKenglish]{babel}
\usepackage{upgreek}
\usepackage{type1cm}
\usepackage{mathptmx}
\usepackage{caption}
\usepackage{subcaption}
\usepackage{siunitx}
\usepackage{textcomp} 
\usepackage{float}
\usepackage{comment}
\usepackage{tikz}
\usetikzlibrary{positioning}
\usepackage{tcolorbox}
\usepackage{multirow}
\usepackage{array}
\usepackage{hyperref}
\newcommand*\diff{\mathop{}\!\mathrm{d}} 
\renewcommand{\baselinestretch}{0.975}
\title{\LARGE \bf
Tissue Activation Calculation in Dual-lead Deep Brain Stimulation* 
}

\author{Anna Franziska Frigge $^{1}$ and Alexander Medvedev $^{1}$
\thanks{*This work is part of the project ``Patient-specific dynamical modeling and optimization of deep brain stimulation" funded within The EU Joint Programme – Neurodegenerative Disease Research 
by the Swedish Research Council, Grant 2020-02901.}
\thanks{$^{1}$ Information Technology,
        Uppsala University, SE-75236 Uppsala, Sweden
        {\tt\small \{anna.frigge,alexander.medvedev\}@it.uu.se}}
}

\begin{document}

\maketitle
\thispagestyle{empty}
\pagestyle{empty}

\begin{abstract}
Deep Brain Stimulation (DBS) is a well-established  neurosurgical treatment aiming at symptom alleviation in a range of neurological and psychiatric diseases. Computational models of DBS are widely used to investigate the effects of stimulation on neural tissue, to explore stimulation targets and sweetspots, and ultimately, to aid clinicians in the DBS programming by calculating the stimulation parameters. Commonly, DBS is performed bilaterally, i.e. with one lead in each brain hemisphere, where computational models are solved independently for one lead at a time. This paper treats scenarios where multiple DBS leads are implanted in close proximity to one another, resulting in interacting electrical fields and, therefore, potentially overlapping stimulation spreads. In particular, a global dual-lead model is compared to approximations derived from single-lead approaches in a cohort of twelve multiple sclerosis (MS) tremor patients. It is concluded that simple superposition of volumes of tissue activated (VTAs) underestimates activation, while superposition of electric fields or activating functions leads to overestimation.
It is concluded that given close proximity of DBS leads, the VTA cannot be computed individually as stimulation fields exhibit significant and complex interaction. The approach is extended to modeling two obsessive compulsive disorder patients with medially placed leads, where similar VTA discrepancies as in the MS patient cohort are observed. 
\end{abstract}

\section{Introduction}
Deep Brain Stimulation (DBS) is a widely practiced neurosurgical treatment method that has had great impact on the management of chronic neurological and psychiatric disorders. Most prominent success examples are  alleviating symptoms in movement disorders such as Parkinson's disease, essential tremor, and dystonia. More recently, DBS therapy has been progressively extended to psychiatric conditions. While currently the only FDA-approved psychiatric condition is treatment-refractory obsessive-compulsive disorder (tr-OCD), DBS has shown promising results in e.g. major depression~\cite{Frey2022}.
Despite decades of clinical application, the underlying mechanisms of DBS remain poorly understood. The complexity of its anatomical targets as well as the variability in therapeutic outcomes and side effects across conditions make DBS an active area of research. Moreover, DBS programming, i.e., the process of identifying optimal stimulation parameters for individual patients, remains challenging due to the increasingly complex lead designs and a large parameter space that can only be sparsely explored in clinical practice~\cite{Patrick2024}.

Computational DBS models have become widespread as versatile tools to investigate the effects of electrical stimulation on neural tissue, to better understand the underlying neural processes, and to aid clinical DBS programming~\cite{Patrick2024,Frigge2024PP}. These models enable researchers and clinicians to predict the spatial distribution of electric fields (EFs), estimate the volume of tissue activated (VTA) by stimulation, and identify which neural structures and pathways are likely to be modulated by a given set of stimulation parameters. In recent years, image-guided and tractography-guided DBS programming~\cite{Torres2024,vanderLinden2025} approaches have gained considerable popularity, offering the promise of more precise, patient-specific targeting, and parameter optimization. 

The overall objective in DBS programming is to maximize therapeutic benefit while minimizing side effects, which can be formulated as an optimization problem. Consider a grid-based scheme where the goal is to minimize the activation metric $\psi$, e.g., EF-norm or activating function (AF), at all points outside the target volume $\Omega_t$, while ensuring sufficient activation within the target. Optimization strategies for DBS programming typically rely on precomputed unit stimulus solutions. Two approaches can be distinguished based on how contact interactions are modeled.

\textbf{Strategy 1:}
Optimal stimulation amplitudes are determined from solutions under a unit input, computed for each contact individually with all others floating, via
\begin{equation}
\begin{aligned}
  \mathbf{u}^* &= \arg \min_{\mathbf{u}} \quad \sum_{\mathbf{r}_{j} \in \Omega_c} \psi(\mathbf{r}_{j}, \mathbf{u}),\\
       \text{s.t.} \quad & \psi(\mathbf{r}_{i}, \mathbf{u}) \geq \psi_{\mathrm{th}} \quad \text{for at least } \theta \cdot n(\Omega_t) \text{ points in } \Omega_t,
\end{aligned}
\label{eq:optimization_direct}
\end{equation}
where $\mathbf{u} \in \mathbb{R}^{N_c}$ represents stimulation amplitudes on $N_c$ contacts and zero amplitude equals an inactive contact. $\mathbf{r}_i \in \Omega_t$ and $\mathbf{r}_j \in \Omega_c$ denote spatial coordinates in the target and complement regions, $\psi_{\mathrm{th}}$ is the activation threshold, $\theta \in [0,1]$ is the minimum target coverage fraction, and $ n(\cdot)$ is the cardinality of the set. This approach uses superposition of independent single-contact solutions, thereby neglecting non-zero boundary conditions on floating  contacts induced by the electric field emitted by the active ones~\cite{Frigge2025EMBC}.

\textbf{Strategy 2:}
The optimal contact configuration and amplitude are determined by evaluating all feasible configurations through
\begin{align}\label{eq:optimization_config}
  (\mathbf{c}^*, \lambda^*) &= \arg \min_{\mathbf{c} \in \mathcal{C}} \quad  f(\mathbf{c}),  \\
  f(\mathbf{c}) &= \min_{\lambda}  \sum_{\mathbf{r}_{j} \in \Omega_c} \psi(\mathbf{r}_{j}, \mathbf{c}, \lambda), \nonumber \\
       \text{s.t.} \quad  \psi(\mathbf{r}_{i}, \mathbf{c}, \lambda) &\geq \psi_{\mathrm{th}} \quad \text{for at least } \theta \cdot n(\Omega_t) \text{ points in } \Omega_t, \nonumber
\end{align}
where $\mathcal{C}$ is the set of all possible contact configurations,  $\mathbf{c} \in \mathcal{C}$ denotes a contact configuration and $\lambda$ the stimulation amplitude. For each $\mathbf{c}$, the inner optimization determines the minimal amplitude satisfying the target coverage constraint, and configurations are ranked by $f(\mathbf{c})$.

Evidently, the performance of such optimization strategies depends on the accuracy of the underlying computational models used to estimate stimulation spread and its interaction with patient-specific anatomy. Typically, simulations of computational DBS models are performed for one lead at a time regardless of how many leads are present under the assumption that multiple leads do not interact or influence each other's stimulation spread. While this approach is computationally efficient and has proven adequate for many clinical scenarios involving bilateral stimulation in separate hemispheres, it may not accurately capture the complex field interactions that occur in certain configurations. In particular, when DBS leads are placed in close proximity, such as in dual-lead DBS~\cite{Oliveria2017,Maesawa2022,Wong2022} where multiple leads are positioned in  close spatial proximity~\cite{Maarouf2016}, the assumption of independence may break down. 

To the authors' best knowledge, this paper investigates for the first time how VTA approximations differ when derived from a global dual-lead model versus combinations of independent single-lead models. The main contributions of this paper include:
\begin{enumerate}
    \item It is demonstrated that, in closely placed DBS leads, unilateral stimulation  exerted by one lead induces currents in conducting contact (floating) surfaces of  inactive  leads, thus violating the independence assumption.
    \item  Systematic biases in VTA approximations are quantified: simple VTA superposition underestimates activation compared to the global dual-lead model, while superposition of underlying physical quantities (combined EF or AF) overestimates activation counts.
    \item  Clinical target coverage is analyzed across methods showing that the biases extend beyond VTA approximations to clinical target coverage metrics.
\end{enumerate}

The rest of this paper is organized as follows: First, the underlying mathematical model of DBS and two different metrics for computing neural activation are presented. Next, the patient data underlying the simulations are described, followed by presentation and discussion of the results.

\section{Mathematical Modeling}
\begin{figure}
    \centering
    \includegraphics[width=0.75\linewidth]{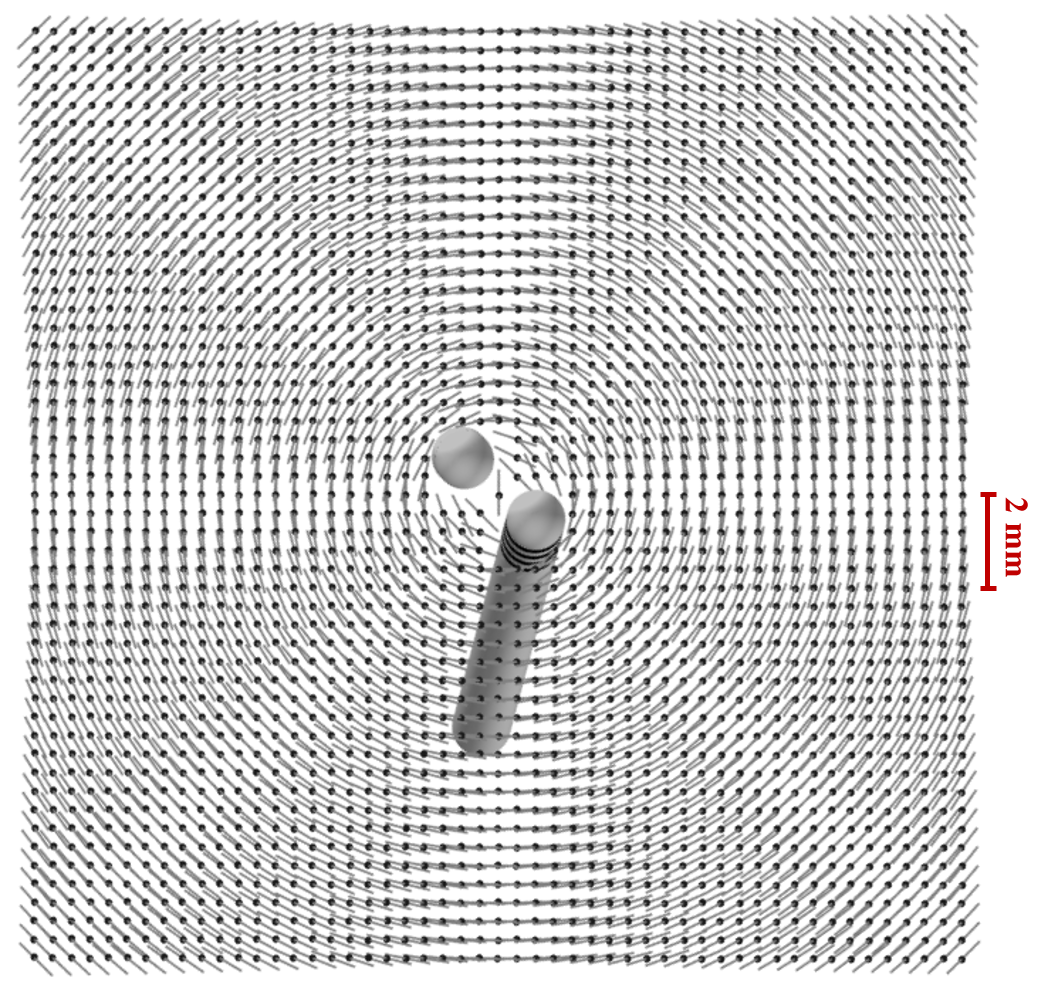}
    \caption{Grid of neurons surrounding the leads. For illustration, only a single plane of the full $20 \times 20 \times 20$~mm uniform grid (node spacing: \SI{0.4}{mm}) is shown. Each axon (light grey) has a length of \SI{1}{mm}. }
    \label{fig:Neuron_grid}
\end{figure}

\subsection{DBS Model}
To simulate the activation of neural tissue under clinical stimulation settings, patient-specific finite element models of the dual-lead DBS were built and simulated in COMSOL Multiphysics\textsuperscript{\textregistered} 6.3 using LiveLink\textsuperscript{\texttrademark} for MATLAB. Two modeling approaches were compared: a single-lead model solving for each lead independently, and a global dual-lead model simulating both leads simultaneously as illustrated in Fig.~\ref{fig:Neuron_grid}. An isotropic conductivity map $\sigma_{\mathrm{iso}}$ was generated from tissue segmentation (gray matter, white matter, and cerebrospinal fluid) of the MNI152 template MRI, using literature-derived conductivity values~\cite{Andreuccetti2017}. 
The anisotropic conductivity tensor $\mathbf{\sigma}$ was derived from the averaged diffusion tensor $\mathbf{D}$ obtained from the IIT Human Brain Atlas~\cite{Zhang2018IIT} using a volume-constraint approach~\cite{Miranda2001}
\begin{equation}
\mathbf{\sigma} = 3 \cdot \frac{\sigma_{\mathrm{iso}}}{\text{tr}(\mathbf{D})} \cdot \mathbf{D},
\end{equation}
where $\text{tr}(\mathbf{D})$ denotes the trace  of the diffusion tensor and $\sigma_{\mathrm{iso}}$ represents the tissue-specific isotropic conductivity value at each voxel.

The conductivity tensor was applied within a $\SI{50}{\milli\meter} \times \SI{50}{\milli\meter} \times \SI{50}{\milli\meter}$ volume centered at either the lowermost contact in the single-lead model or the midpoint between the two lowermost contacts in the dual-lead model. Beyond this region, tissue conductivity was modeled as homogeneous. 

For the inhomogeneous tissue box, a fine mesh with a maximum element size of $\SI{5}{mm}$ was applied, while a coarser mesh element size was implemented further away from the DBS leads.
The single lead and the double lead models included a total of $965,679$ and $1,869,385$ mesh elements, respectively. 

The stimulation scenario was modeled through the partial differential equation 
\begin{equation}
    \nabla \cdot (\sigma \nabla u) = 0,
    \label{eq:pde_static}
\end{equation}
where $\nabla \cdot$ is the divergence operator,  $\sigma$ denotes the conductivity, and $\nabla u$ stands for the gradient of the electric potential. All non-active contact surfaces were given a floating boundary condition given by the surface integral
\begin{equation}
    \int \limits_{\partial\Omega} \mathbf{J}\cdot \mathbf{n}~\diff S = 0,
    \label{eq:floating_boundary}
\end{equation}
where $\mathbf{J}$ represents the current density, $\mathbf{n}$ is the normal vector to the contact surface.
Since the patients were implanted with a constant-voltage system, the boundary condition on the active contacts was set to a constant potential $u=u_0$, where $u_0$ is each patient's individual stimulation amplitude. The quasi-static model in~\eqref{eq:pde_static} is useful for characterizing the spatial extent of neural activation but does not model temporal dynamics such as pulse width modulation, stimulation frequency effects, or phase-dependent interactions in multi-lead configurations.
Bipolar settings were modeled with active cathodic contacts and the anodic contacts set as the return path (ground).

\subsection{Volume of Tissue Activated}
Both the single- and the dual-lead models were evaluated on the same $20 \times 20 \times 20$ mm ($30 \times 30 \times 30$ mm for the cohort from Maarouf et al.~\cite{Maarouf2016}) grid with a grid spacing of $\SI{0.4}{mm}$ between nodes. The set of grid points is denoted as $\Omega$. For each patient the grid was centered at the midpoint between the lowermost contacts of both leads. A 2D projection of the grid is illustrated in Fig.~\ref{fig:Neuron_grid}.

Here, we consider two approaches to quantify the VTA:
\begin{enumerate}
    \item \textbf{EF-norm} -- The simplest and most widely used approach identifies activated points as those locations $\mathbf{r}$ where the EF-norm $||\mathbf{E}||$ exceeds a given threshold $E_{\text{th}}$,  i.e.
    \begin{equation}
        ||\mathbf{E}(\mathbf{r})|| = \sqrt{E_x^2 + E_y^2 + E_z^2} \geq E_{\mathrm{th}},
        \label{eq:Enorm}
    \end{equation}
    where $\mathbf{E}=-\nabla u$ is the EF derived from the electric potential $u$, and $E_x$, $E_y$, and $E_z$ denote its cartesian components. Pulse-width adjusted activation thresholds were derived from Åstr{\"o}m et al.~\cite{Astrom2015}.
\item \textbf{AF-Max} -- A more sophisticated approach based on the second spatial derivative of the electric potential along neuron trajectories, as described by Duffley et al.~\cite{Duffley2019}. The tangential component of the AF along an axon with tangent vector $\mathbf{t}\in \mathbb{R}^3$ is given by
\begin{equation} \varphi_{\mathrm{tan}}(s) = -\frac{\partial^2 u}{\partial s^2} = -\mathbf{t}^\intercal \cdot \mathbf{H} \cdot \mathbf{t}.
\end{equation}
where $\mathbf{H}$ is the Hessian matrix of the electric potential and $s$ is the arc length along the axon. 
Each axon trajectory was centered at the respective grid point and the axon length of \SI{1}{mm} was sampled every $\SI{0.1}{mm}$. To account for maximum excitability, each axon was oriented perpendicular to the average trajectory of the two DBS leads. Both the grid points and axon trajectories are illustrated in Fig.~\ref{fig:Neuron_grid}. For each trajectory, the maximum absolute value along its length is computed via 
\begin{equation}
\varphi_{\mathrm{max}} = \max_s |\varphi_{\text{tan}}(s)|.
\end{equation}
A point is considered activated if $\varphi_{\mathrm{max}} \geq \varphi_{\text{th}}$, where pulse-width adjusted thresholds are again applied based on simulation results from Duffley et al.~\cite{Duffley2019}. This approach better captures the spatial gradient effects and is supposedly better suited for modeling bipolar stimulation~\cite{Duffley2019}. 

\end{enumerate}
Available software tools like lead-DBS~\cite{Neudorfer2023} typically compute VTAs from independent computational models for each lead.
Consequently, end users are likely to only have access to the individual VTAs computed separately for each lead. From these individual VTAs, the dual-lead VTA can be approximated by simple superposition of the activation regions, given by
\begin{equation}
\begin{aligned}
    V_{\mathrm{EF}} &= \{\mathbf{r} \in \Omega : \max(||\mathbf{E_1}(\mathbf{r})||, ||\mathbf{E_2}(\mathbf{r})||) \geq E_{\mathrm{th}}\}, \quad \text{and}\\
    V_{\mathrm{AF}} &= \{\mathbf{r} \in \Omega : \max(\varphi_{\mathrm{max,1}}(\mathbf{r}), \varphi_{\mathrm{max,2}}(\mathbf{r})) \geq \varphi_{\mathrm{th}}\},
\end{aligned}
\label{eq:VTAsuperimposed}
\end{equation}
for the EF-norm and AF-Max approaches, respectively.
However, following directly from the linearity of~\eqref{eq:pde_static}, another approximation of the dual-lead scenario can be obtained from these independent solutions by first superimposing the underlying physical quantities, specifically, the EF vectors and Hessian matrices, before applying activation thresholds.
For the EF-norm approach, the combined activation region is defined as
\begin{equation}
    C_{\mathrm{EF}} = \{\mathbf{r} \in \Omega : ||\mathbf{E_1}(\mathbf{r}) + \mathbf{E_2}(\mathbf{r})|| \geq E_{\mathrm{th}}\},
    \label{eq:Enorm_combined}
\end{equation}
where $\mathbf{E_1}$ and $\mathbf{E_2}$ are the EF vectors from the individual lead models.
For the AF approach, the combined activation region is obtained by first summing the AFs along each trajectory before taking the maximum, i.e. 
\begin{equation}
\begin{aligned}
    C_{\mathrm{AF}} = \Bigg\{\mathbf{r} \in \Omega : \max_s |\varphi_{\text{tan,1}}(s) +\varphi_{\text{tan,2}}(s)| & \geq \varphi_{\text{th}}\Bigg\} \\
    = \Bigg\{\mathbf{r} \in \Omega : \max_s |\mathbf{t}^\intercal  (\mathbf{H_1}+ \mathbf{H_2}) \mathbf{t}| & \geq \varphi_{\text{th}}\Bigg\},
 \end{aligned}
\label{eq:AFmax_combined}
\end{equation}
where $\mathbf{H_1}$ and $\mathbf{H_2}$ are the Hessian matrices of the electric potential from each lead model.

Notably, the independent solution cannot capture the influence of induced currents at inactive (floating) contacts at the other lead. Hence, in the following, the approximations from the independent models are compared to the global solution directly obtained from the dual-lead model.
\label{sec:approximations}

\begin{figure}
    \centering
    \begin{subfigure}[t]{0.4\linewidth}
      \includegraphics[height=4.2cm]{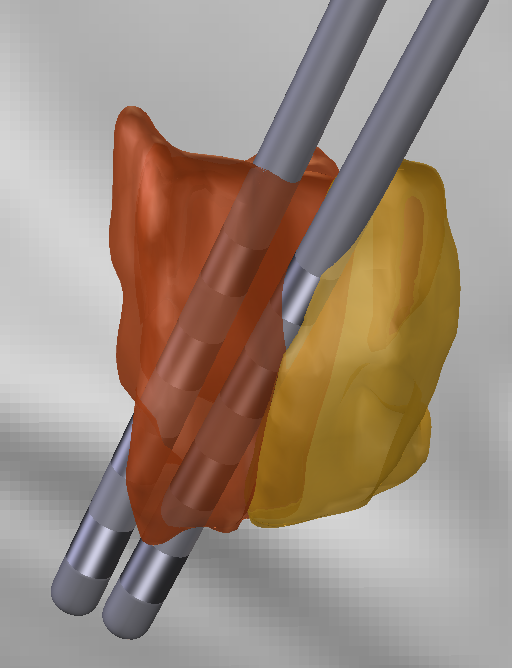}
      \caption{}
\end{subfigure}
\begin{subfigure}[t]{0.58\linewidth}
      \includegraphics[height=4.2cm]{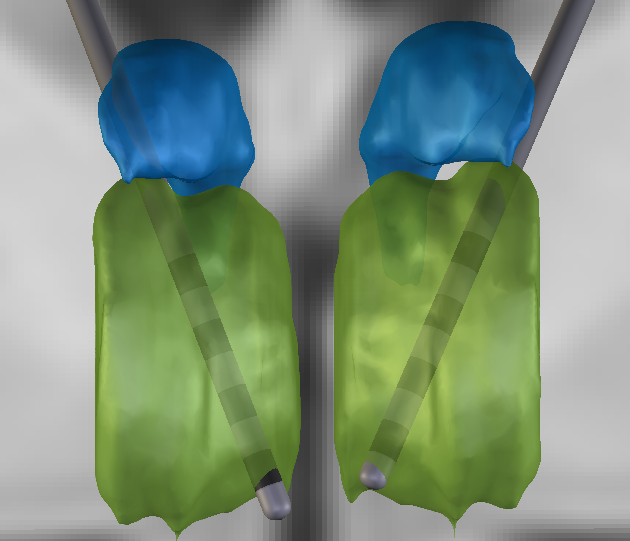}
      \caption{}
\end{subfigure}
    \caption{Lead placement for (a) one patient from cohort I relative to the VIM (red) and the VO (orange)~\cite{Ewert2018} and (b) one patient from cohort II relative to the MD (green) and the anterior (blue) nuclei of the thalamus\cite{Ilinsky2018}. The figures were rendered in Lead-DBS~\cite{Neudorfer2023}. }
    \label{fig:Reconstructions}
\end{figure}
\section{Simulation}
\begin{figure*}
    \centering
    \begin{subfigure}{0.32\textwidth}
        \includegraphics[width=\linewidth]{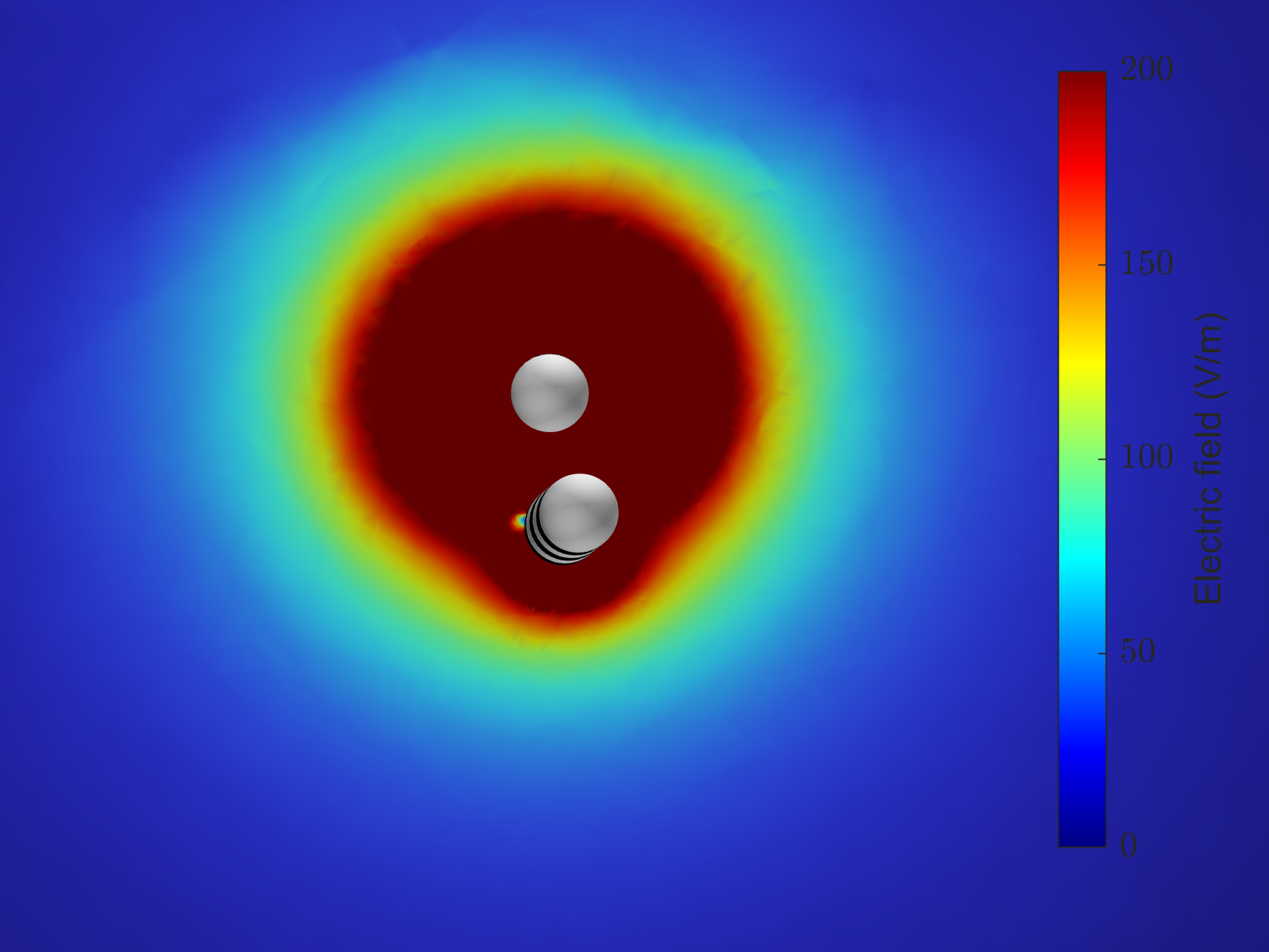}
        \caption{ViM lead active}
        \label{fig:image1}
    \end{subfigure}
    \hfill
    \begin{subfigure}{0.32\textwidth}
        \includegraphics[width=\linewidth]{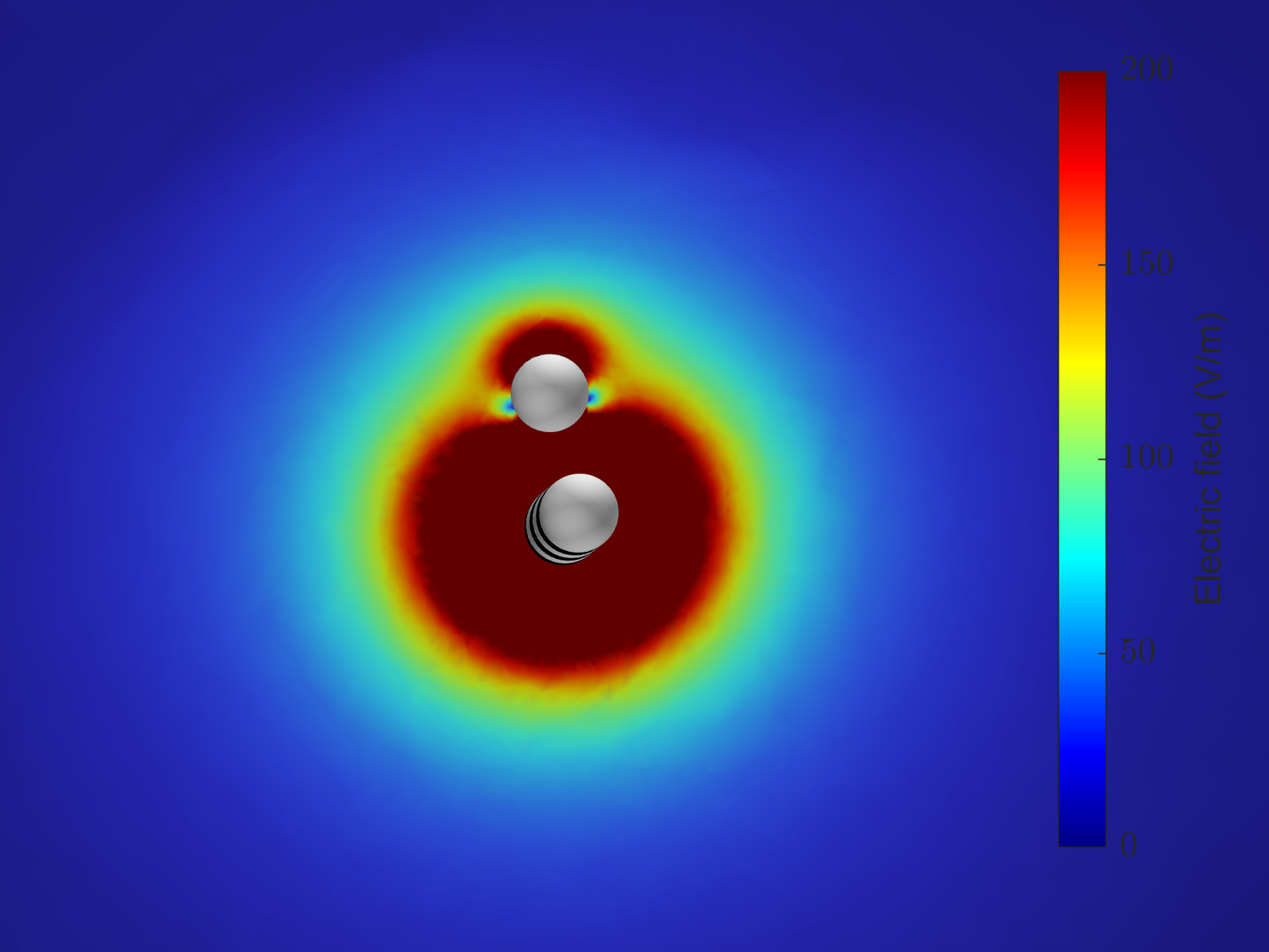}
        \caption{VO lead active }
        \label{fig:image2}
    \end{subfigure}
    \hfill
    \begin{subfigure}{0.32\textwidth}
        \includegraphics[width=\linewidth]{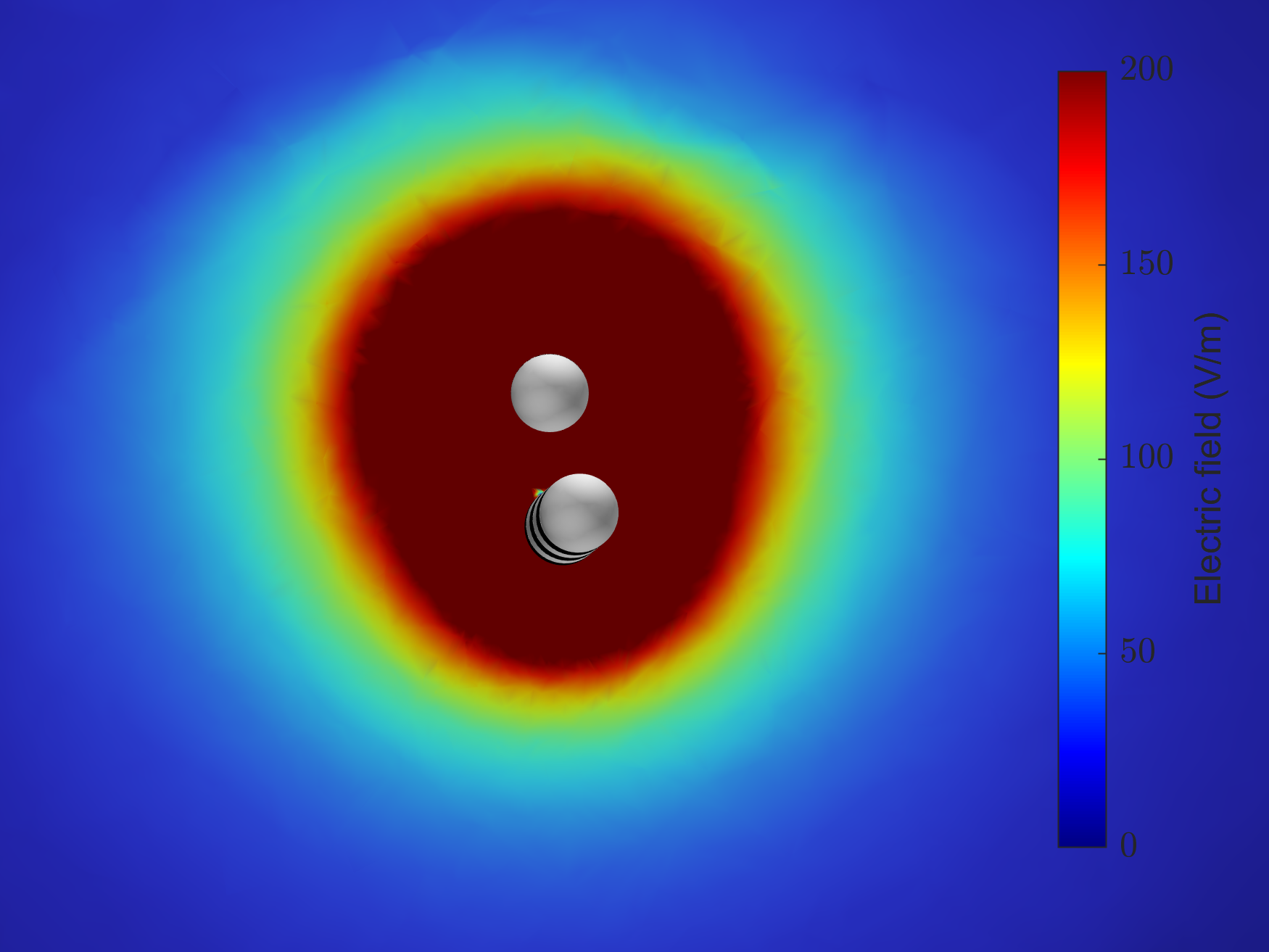}
        \caption{Both leads active}
        \label{fig:image3}
    \end{subfigure}
    \caption{VTA computed from \eqref{eq:Enorm} for the two DBS leads in patient 16. VTAs were computed under clinical settings with a) only the ViM lead, b) only the VO lead, and c) both leads active. A clear effect of the floating contacts on the non-active lead can be observed.}
    \label{fig:VTA_dual}
\end{figure*}
The simulations presented in this study are based on DBS lead coordinates of two patient cohorts previously reported by Oliveria et al. (2017)~\cite{Oliveria2017} and Maarouf et al. (2016)~\cite{Maarouf2016}. The first cohort~\cite{Oliveria2017}, referred to as Cohort~I in the following, comprises 12 individuals with multiple sclerosis tremor. 
Each patient was implanted with two Medtronic 3387 DBS leads in the hemisphere contralateral to their most symptomatic upper extremity. One electrode was placed at the border of the ventralis intermedius (VIM) and ventralis oralis posterior (VOP) nuclei - referred to hereafter as the VIM lead - and a second electrode at the ventralis oralis anterior (VOA)-VOP border - referred to as the VO lead. 

The second cohort~\cite{Maarouf2016}, referred to as Cohort~II in this manuscript, includes four patients with therapy-refractory obsessive compulsive diroser (tr-OCD), however only the two of patients (ID 2 and 4) that were also implanted with Medtronic 3387 DBS leads were considered in this study. The leads were placed in the medial dorsal (MD) and the ventral anterior (VA) nucleus of the thalamus. 

The original coordinates from~\cite{Oliveria2017} and~\cite{Maarouf2016} were specified relative to the mid-commissural point (MCP) and the anterior-comissural point (AC) of the individual patient, respectively. In the present work, these coordinates were transformed into the ICBM 2009b nonlinear asymmetric MNI space. 
The lead placement for one patient from~\cite{Oliveria2017} and one patient from~\cite{Maarouf2016} relative to the respective target structures are displayed in Fig.~\ref{fig:Reconstructions}.

To quantify differences in VTAs beyond volumetric comparison, target coverage was evaluated using a grid-based approach. Target volumes were mapped onto a 3D grid, and grid points were divided into two disjoint sets: a target region $\Omega_t$ and its complement $\Omega_c$ (i.e., all grid points outside the target), i.e. $\Omega = \Omega_t \cup \Omega_c$. The percentage of target coverage was estimated by summing activated grid points within $\Omega_t$. To enable cross-cohort comparison, all coverages were normalized by the target percentage covered by the respective dual model. 

For Cohort~I, coverage of the ViM and VO was evaluated based on the Distal atlas~\cite{Ewert2018}. Similarly, for Cohort~II, the MD and anterior nucleus of the thalamus were evaluated using the Human Motor Thalamus atlas~\cite{Ilinsky2018}.


\section{Results}
This section presents results from the dual-lead simulations and their single-lead approximations.
\subsection{Dual lead model}
In the clinical settings, both leads were active in all patients. However, to illustrate the effect of induced currents from one lead on the another lead, the dual-lead model was first solved with each lead activated individually and then with both leads active simultaneously. The results for one patient are shown in Fig.~\ref{fig:VTA_dual}, where activation of a single lead induces charge accumulation on the floating contacts of the inactive lead. 

\subsection{Approximation from single lead models}
\begin{figure}
    \centering
    \includegraphics[width=\linewidth]{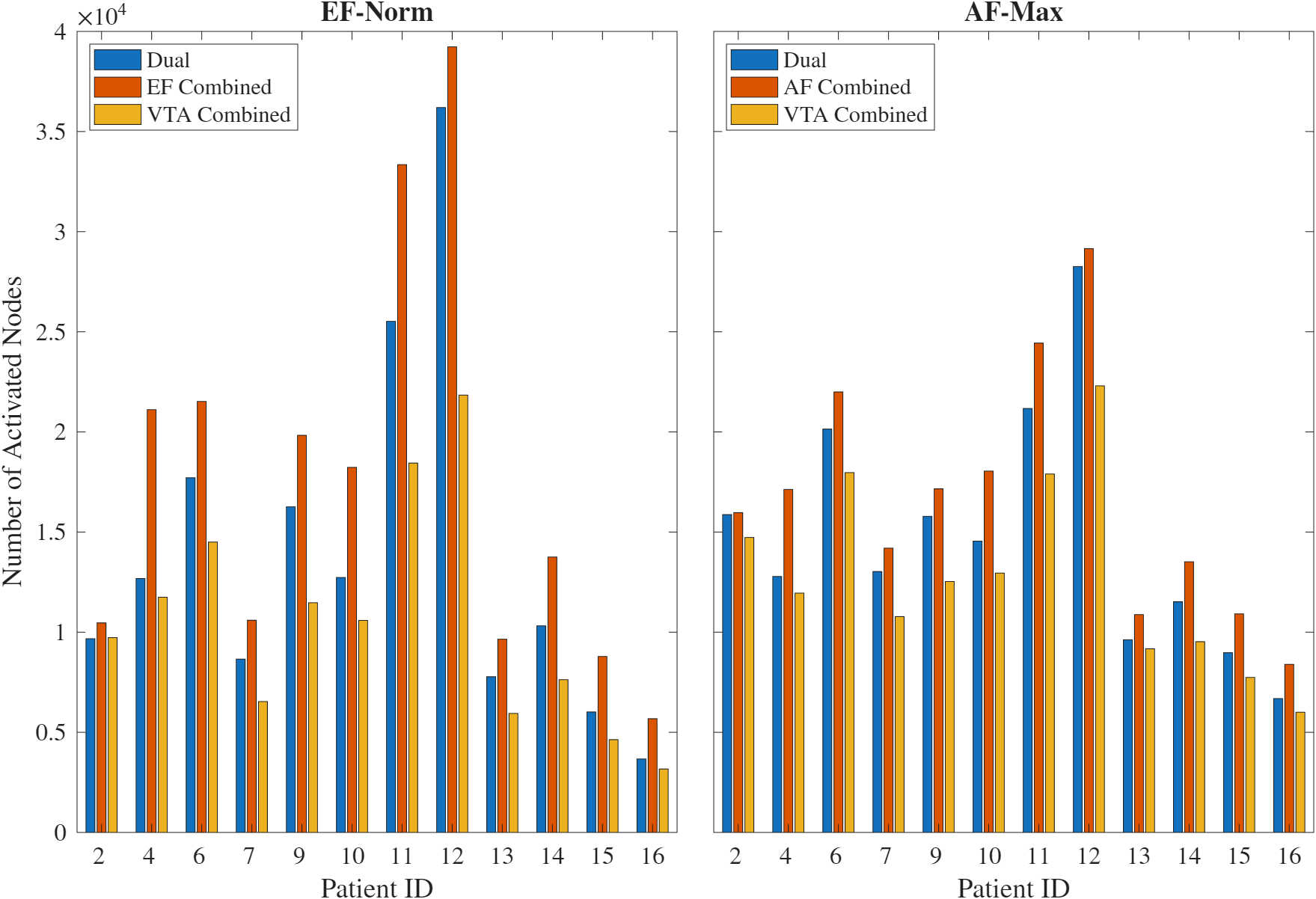}
    \caption{Number of activated nodes in Cohort~I for the EF-norm (left) and AF Max (right), respectively. The dual solution is approximated by the two approaches presented in ~\ref{sec:approximations}.}
    \label{fig:Cohort_TotalActivations}
\end{figure}
The dual-lead solution for all patients in Cohort~I was approximated using the two approaches described in Section~\ref{sec:approximations}: (i) superposition of VTAs and (ii) superposition of the underlying physical quantity (EF or AF/Hessian) before applying the activation threshold. Fig.~\ref{fig:Cohort_TotalActivations} shows the total number of activated grid nodes under clinical stimulation settings, where patient IDs correspond to those in the original publication~\cite{Oliveria2017}. The total number of activations varies considerably across patients due to differences in stimulation amplitudes, pulse widths, and lead positions. For cohort-level comparison, the number of activations for each patient were normalized to the total number of activations in the respective dual model ($=100 \%$) in subsequent analysis. 

\begin{figure}
    \centering
    \includegraphics[width=\linewidth]{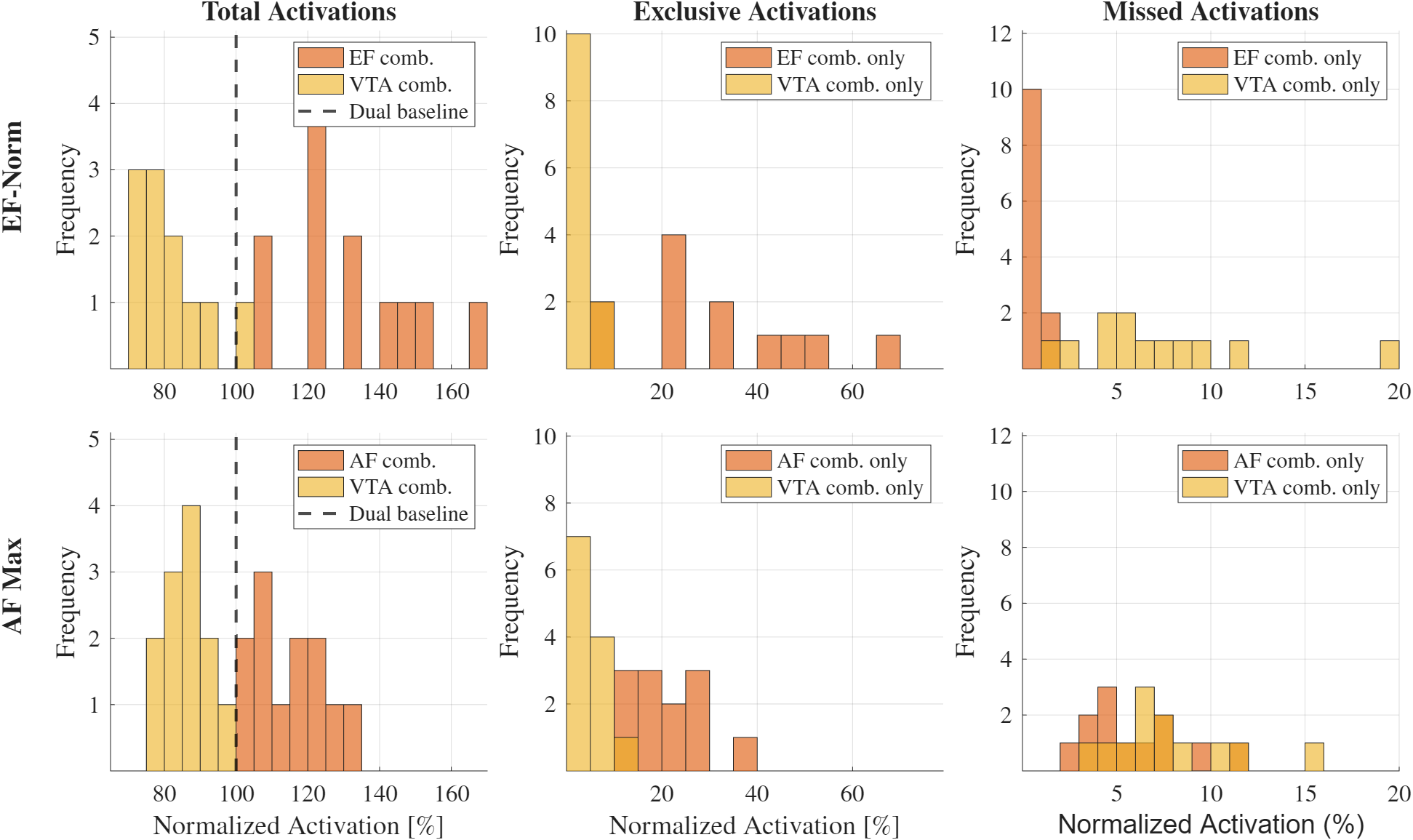}
    \caption{Distribution plots of total activations, exclusive activations, and missed activations for cohort I. The number of activations were normalized with respect to the total number of activations in the respective dual model.}
    \label{fig:cohort_activation_distributions}
\end{figure}

Fig.~\ref{fig:cohort_activation_distributions} shows cohort-level distributions of total activations, exclusive activations (i.e., overestimated activations relative to the dual model), and missed activations (i.e., underestimated activations relative to the dual model).
For both the EF-norm and AF-Max approach simple VTA combination tends to underestimate the total number of activations relative to the dual model, whereas the combined EF and AF methods tend to overestimate activation counts. 

\begin{figure}
    \centering
    \includegraphics[width=\linewidth]{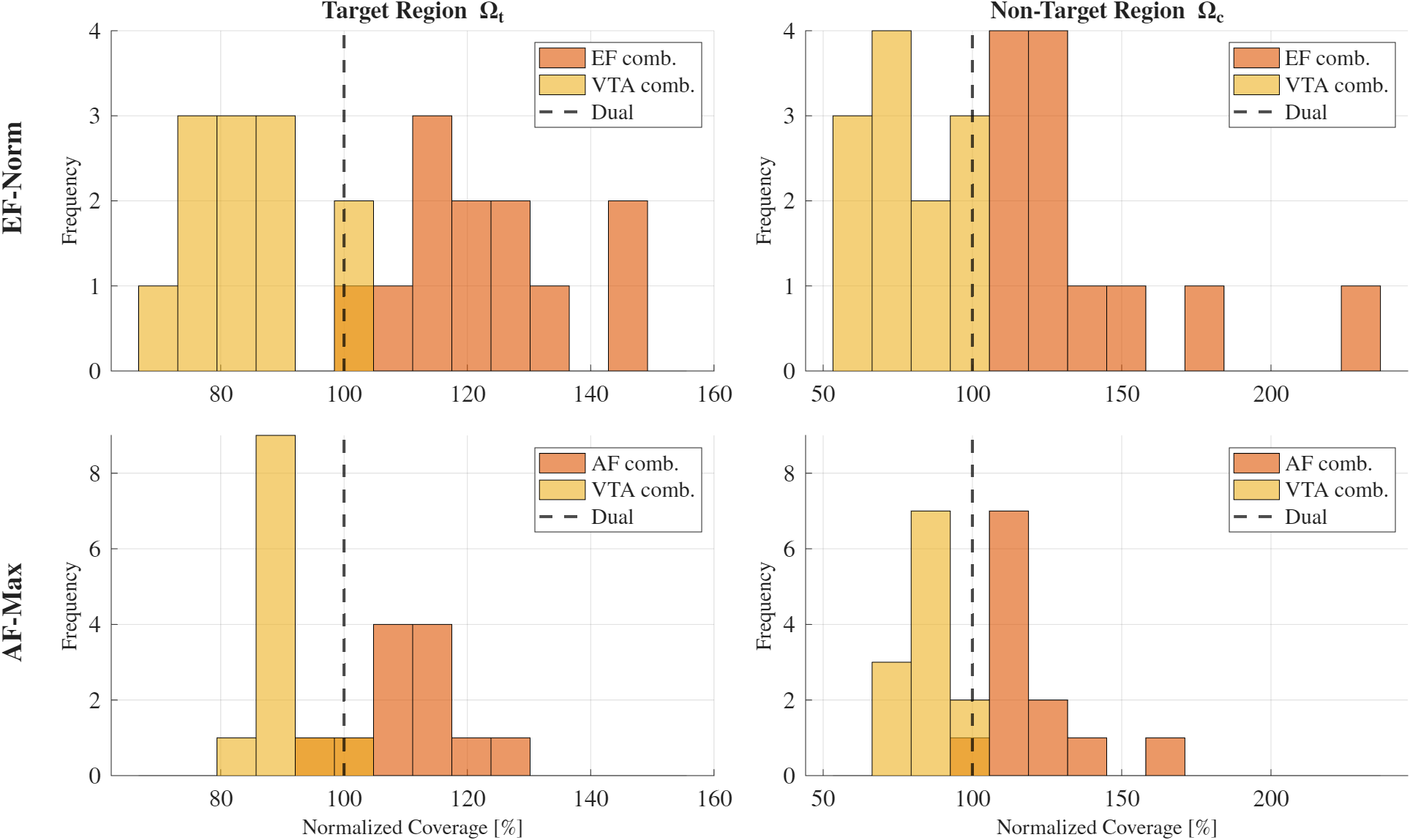}
    \caption{Distribution plots of target and non-target coverages for Cohort~I. For each patient the target and non-target coverages were normalized with respect to the coverages in the respective dual model.}
    \label{fig:cohort_target_coverages}
\end{figure}

Similar effects can be observed when explicitly evaluating coverages of target and non-target regions, shown in Fig.~\ref{fig:cohort_target_coverages}.

\subsection{Bilateral dual-lead DBS}
\begin{figure}
    \centering
    \footnotesize{\textbf{Patient 2}}\\
    \includegraphics[width=\linewidth]{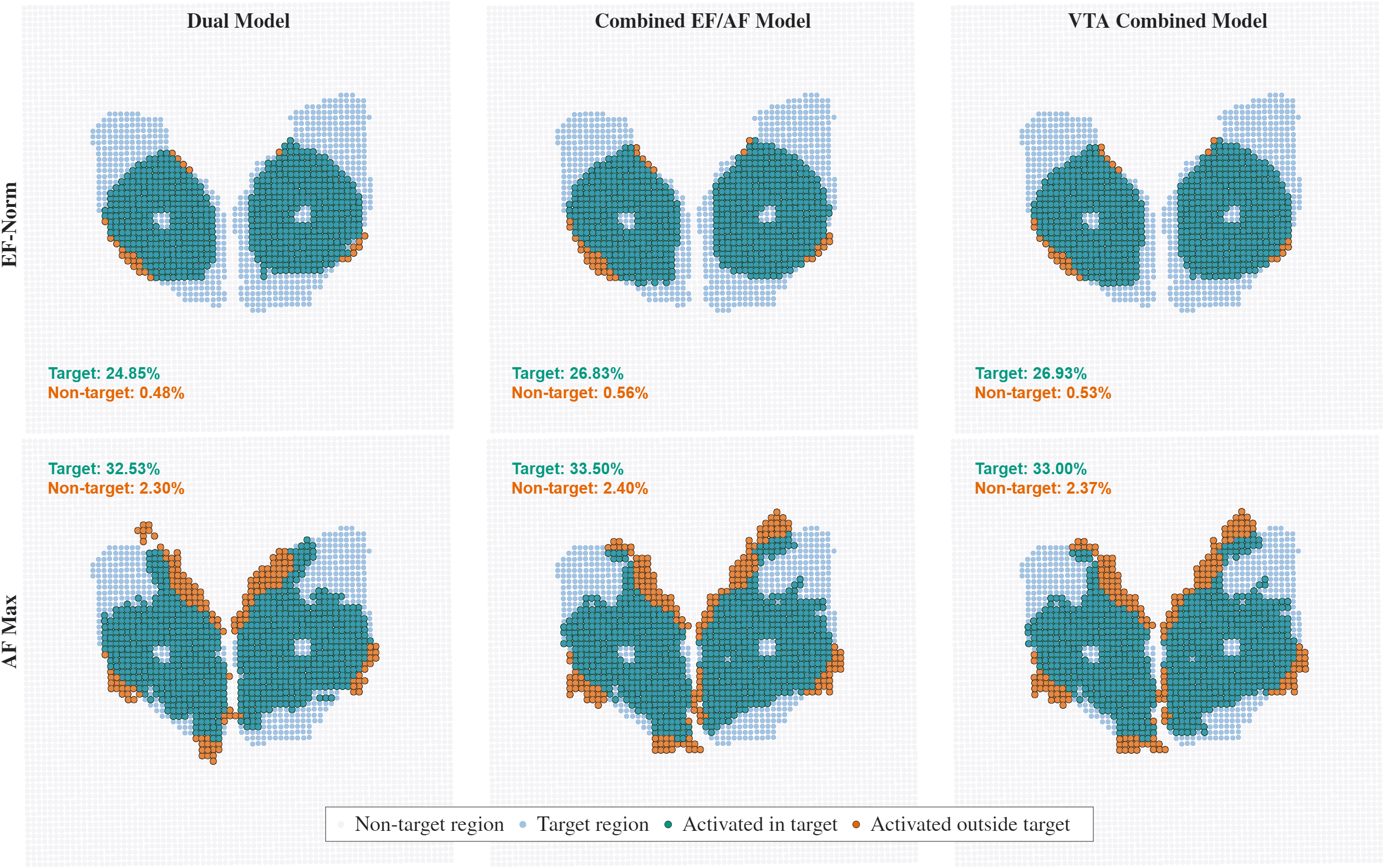}
    \footnotesize{\textbf{Patient 4}}\\
    \includegraphics[width=\linewidth]{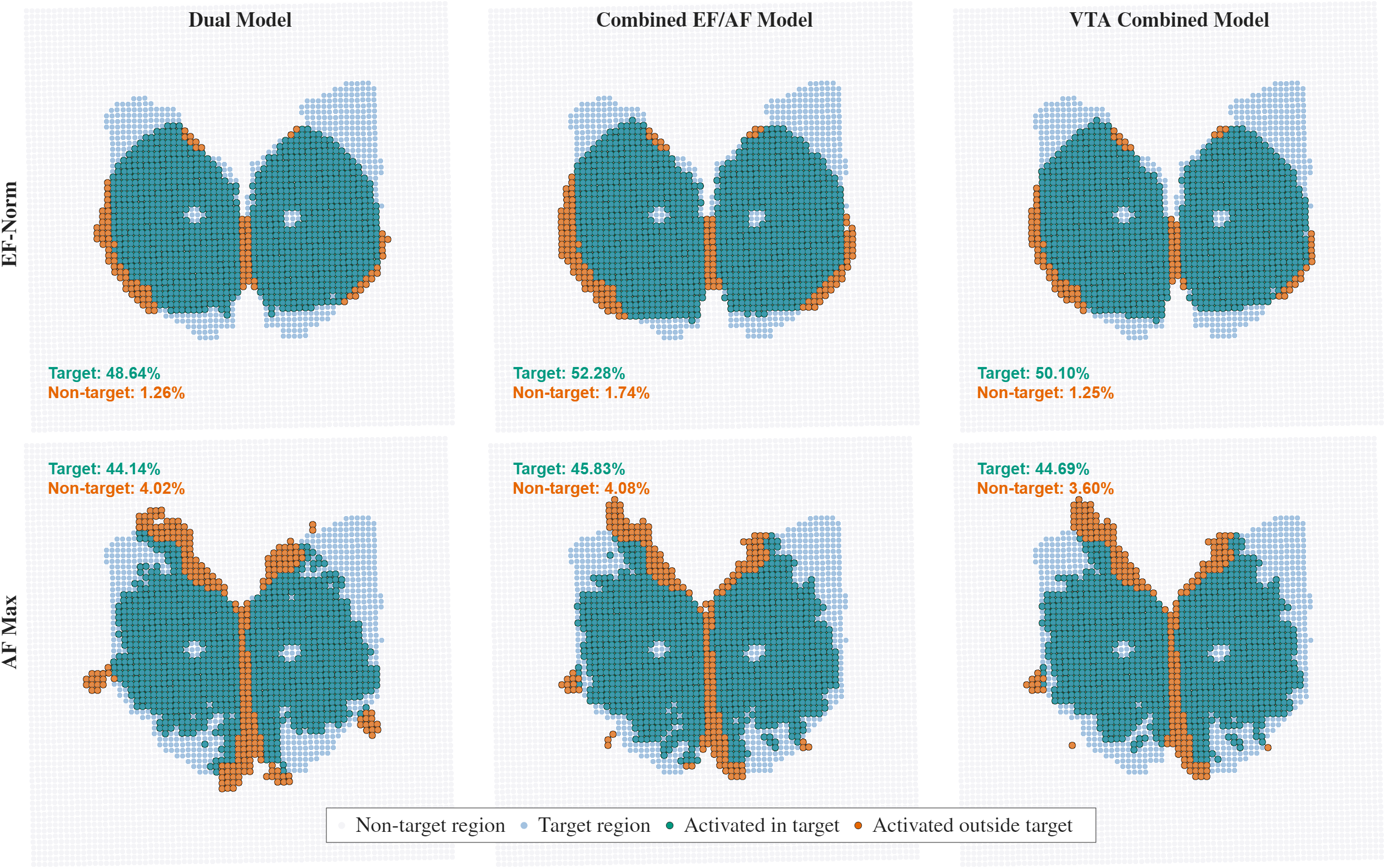}
    \caption{A 2D projection of the target coverages for both patients in Cohort~II. Patient IDs correspond to those used in the original publication. For each patient and model, the 3D target and non-target activations for each model and patient are given in percent of $n(\Omega_t)$ and $n(\Omega_c)$, respectively. With over $57,000$ nodes in $\Omega_t$ and over $370,000$ nodes in $\Omega_c$, even small percentages make up for large differences in the number of activated nodes.
 }
    \label{fig:maarouf_target_coverages}
\end{figure}

Results for the two patients in Cohort~II are shown in Fig.~\ref{fig:maarouf_target_coverages}. Unlike the first cohort, where leads were implanted unilaterally, this dataset involves bilateral DBS targeting medial structures. Consequently, the inter-lead distance is greater than in the previous example. Nevertheless, differences in the activation percentages can still be observed. It should be noted that the target and non-target activations are given in percent of  $n(\Omega_t)$ and $n(\Omega_c)$, respectively. With over $57,000$ nodes in $\Omega_t$ and over $370,000$ nodes in $\Omega_c$, even small percentages make up for large differences in the number of activated nodes.


\section{Discussion}
This paper investigates the effects of dual-lead stimulation in scenarios where DBS electrodes are positioned in close proximity.  Despite clinical implementation of dual-lead DBS, computational analysis of stimulation spread interactions in these configurations is scarce in the literature. The results demonstrate that when leads are implanted near each other, their electric fields and activating functions interact in ways that cannot be accurately captured by independent single-lead models; instead, a unified computational model incorporating both leads simultaneously is required. 

While the primary cohort examined here represents an extreme case of two leads implanted within the same hemisphere, other studies have reported similar ipsilateral dual-lead placements in essential tremor~\cite{Chandra2025}, Holmes tremor~\cite{Kobayashi2014}, and complex ipsilateral dystonia~\cite{Chang2022}.
Furthermore, the analysis of two patients with bilateral DBS in Cohort~II demonstrated that dual-lead modeling is also relevant for medially placed electrodes, such as those targeting certain structures in tr-OCD treatment. 

The clinical importance of accurate dual-lead modeling is further supported by a recent study from Chandra et al. (2025)~\cite{Chandra2025}, which reported improved coverage of fibers connected to the supplementary motor area, leading to enhanced tremor control in refractory essential tremor patients. However, their analysis likely employed single-lead simulations, potentially underestimating the contribution of field interactions between bilateral leads to the observed therapeutic outcomes. 

These results have direct implications for model-based DBS programming optimization that use target and non-target coverages for predictions of contact configurations~\cite{Frigge2024PP,Torres2024}. The optimization problems in~\eqref{eq:optimization_direct} and~\eqref{eq:optimization_config} can be solved either independently for each lead or from a global dual-lead model. In configurations with closely positioned leads, the findings suggest that optimization strategies likely need to employ global models that account for field interactions rather than treating leads independently.
Future work should investigate whether the choice of modeling approach substantially affects optimal contact configurations and whether such differences manifest in clinically observable outcomes, thereby informing when global models are necessary versus when independent optimization suffices.

For single forward solutions, the dual-lead model is slightly more computationally efficient than solving two independent single-lead models sequentially (\SI{4}{min} \SI{24}{s} vs. \SI{4}{min} \SI{50}{s} for two single-lead solutions). However, this largely vanishes in optimization contexts. In strategy~\eqref{eq:optimization_direct}, both approaches require $N_c$ forward solutions, i.e. one per contact with all others floating. In strategy~\eqref{eq:optimization_config}, the number of required solutions scales with $n(\mathcal{C})$, the number of feasible contact configurations. For dual-lead systems, $n(\mathcal{C})$ grows combinatorially with contacts across both leads, exceeding the configuration space of independent single-lead optimizations. 

While the the dual-lead model is computationally more efficient than solving two single-lead models sequentially (7.8 min versus 10.7 min), this advantage diminishes in optimization contexts. Both optimization strategies in~\eqref{eq:optimization_direct} and~\eqref{eq:optimization_config} require solving the model repeatedly for different active contact configurations, with the total computational cost dominated by the number of optimization iterations rather than the per-solution modeling approach.

This work has several limitations. All simulations were performed in MNI space, enabling cross-cohort comparisons but neglecting patient-specific anatomy. Additionally, our static modeling approach captures only a temporal "snapshot" of stimulation. Although activation thresholds were pulse-width adjusted, lead interactions likely depend on the temporal synchrony of stimulation pulses. Synchronous versus asynchronous delivery, along with variations in pulse width, frequency, and waveform characteristics, may significantly affect field interactions and activation patterns. Future studies should incorporate biophysically detailed neuron models to capture these temporal dynamics in dual-lead configurations.

\section{Conclusions}
This work demonstrates that closely positioned DBS leads exhibit field interactions that cannot be adequately captured by independent single-lead models, resulting in systematic approximation errors in VTA predictions and target coverage metrics. Hence, model-based programming optimization in dual-lead configurations likely requires global computational models to ensure accurate contact configuration selection, particularly when leads are placed within the same hemisphere or target medial structures where bilateral leads are in close spatial proximity.

\addtolength{\textheight}{-12cm}   






\bibliographystyle{IEEEtran}
\bibliography{references}

\end{document}